\documentstyle[prd,aps,epsfig,tighten,floats,twocolumn]{revtex}

\newcommand{\app}[3]{Astropart.\ Phys.\ {\bf #1}, #3 (#2)}

\newcommand{\hepph}[2]{{\tt hep-ph/#1} (#2)}
\newcommand{\astroph}[2]{{\tt astro-ph/#1} (#2)}
\newcommand{\prep}[3]{Phys.\ Rep.\ {\bf #1}, #3 (#2)}
\newcommand{\plb}[3]{Phys.\ Lett.\ {\bf B#1}, #3 (#2)}

\newcommand{\npb}[3]{Nucl.\ Phys.\ {\bf B#1}, #3 (#2)}
\newcommand{\cpc}[3]{Comm.\ Phys.\ Comm.\ {\bf #1}, #3 (#2)}
\renewcommand{\apj}[3]{Astrophys.\ J.\ {\bf #1}, #3 (#2)}
\renewcommand{\prl}[3]{Phys.\ Rev.\ Lett. {\bf #1}, #3 (#2)}
\renewcommand{\prd}[3]{Phys.\ Rev.\ {\bf D#1}, #3 (#2)}
\renewcommand{\rmp}[3]{Rev.\ Mod.\ Phys.\ {\bf #1}, #3 (#2)}
\newcommand{\jhep}[3]{JHEP {\bf #1}, #3 (#2)}
\newcommand{\mnras}[3]{Mon.\ Not.\ R.\ Astron.\ Soc.\ {\bf #1}, #3 (#2)}

\begin{document}


\wideabs{
\title{%
\null
\vskip-6pt \hfill {\rm\small CU-TP-1032} \\
\vskip-6pt \hfill {\rm\small CWRU-P14-01} \\
\vskip-6pt \hfill {\rm\small MCTP-01-43} \\
\vskip-9pt~\\
The cosmic ray positron excess and neutralino dark matter}

\author{Edward A. Baltz}

\address{ISCAP, Columbia Astrophysics Laboratory, 550 W 120th St., Mail Code
5247, New York, NY 10027, USA \\ E-mail: {\tt eabaltz@physics.columbia.edu}}

\author{Joakim Edsj\"o}

\address{Department of Physics, Stockholm University, SCFAB,
SE-106~91~Stockholm, Sweden \\ E-mail: {\tt edsjo@physto.se}}

\author{Katherine Freese}

\address{Michigan Center for Theoretical Physics, Physics Department,
University of Michigan, Ann Arbor, MI 48109, USA \\ E-mail: {\tt
ktfreese@umich.edu}}

\author{Paolo Gondolo}

\address{Dept.\ of Physics, Case Western Reserve University, 10900 Euclid Ave.,
Cleveland, OH 44106-7079, USA \\ E-mail: {\tt pxg26@po.cwru.edu}}

\maketitle

\vspace{.3in}
\begin{abstract}
Using a new instrument, the HEAT collaboration has confirmed the excess of
cosmic ray positrons that they first detected in 1994.  We explore the
possibility that this excess is due to the annihilation of neutralino dark
matter in the galactic halo. We confirm that neutralino annihilation can
produce enough positrons to make up the measured excess only if there is an
additional enhancement to the signal.  We quantify the `boost factor' that is
required in the signal for various models in the Minimal Supersymmetric
Standard Model parameter space, and study the dependence on various parameters.
We find models with a boost factor $\geq 30$.
Such an enhancement in the signal could arise if we live in a clumpy halo. We
discuss what part of supersymmetric parameter space is favored (in that it
gives the largest positron signal), and the consequences for other direct and
indirect searches of supersymmetric dark matter.
\end{abstract}

}


\section{Introduction}

Several years ago the HEAT collaboration reported an excess of cosmic ray
positrons with energies $\sim$ 10 GeV \cite{heatfrac}.  In the past year they
again measured this excess using a new instrument, and found excellent
agreement \cite{newdata}.  In this paper we revisit the possibility that this
excess is due to annihilations of Weakly Interacting Massive Particles (WIMPs)
in the galactic halo, in particular the neutralinos in supersymmetric models.
The possibility of detecting positrons as annihilation products of WIMPs in the
galactic halo has been discussed previously, both as a continuum and as a
monochromatic source \cite{epcont,kamturner,epline,heatprimary,be,ms99}.  We
update the results of \cite{be} including the new data, and confirm the
conclusions of that paper, finding that neutralino annihilation can not produce
enough positrons to make up the measured excess without an additional
enhancement to the signal.  Recently, this point has been reiterated in
Ref.~\cite{kanepositron}.

In calculating the observed positron flux from annihilations in the halo, we
encounter several astrophysical uncertainties.  First, cosmic ray propagation
is not perfectly understood, though the errors are unlikely to be larger than a
factor of two.  More importantly, the structure of the galactic dark halo is
unknown.  Any clumpiness in the halo serves to enhance the signal, whether it
is a single nearby clump (or one containing the Earth), or a uniform
distribution of clumps.  There is no compelling argument for any particular
value of the enhancement factor, be it unity or in the thousands or more.  In
this paper we carefully discuss the possibility that a clumped galactic halo
could account for the measured positron excess.

We define $B_s$ to be the boost factor that the WIMP annihilation signal from a
smooth galactic halo must be multiplied by to match the HEAT data.  We have
explored models in the Minimal Supersymmetric Standard Model (MSSM) parameter
space to find how large the boost factor must be for each of the models. The
lowest boost factor we found is roughly $30$ for a WIMP that is primarily a
Bino in content with mass $160$ GeV.  For $B_s<100$, we find that the models
are gaugino--dominated, though some have significant Higgsino fractions.  The
masses of the models are in the range 150--400 GeV for the most part.  For
$100<B_s<1000$, the masses are as large as 2 TeV, and some very pure Higgsinos
become allowed.  For both cases there are a significant number of models which
have a large contribution to $a_\mu$ so that the anomalous magnetic moment of
the muon can be explained. We have investigated the dependence of the boost
factor on various parameters.  There is essentially no dependence on parameters
such as $\tan \beta$ or $m_0$ (the sfermion mass scale).  The boost factor does
depend strongly on the relic density of WIMPs; the lowest boost factors are
required for the models with the smallest relic density without rescaling
(defined below).

One can ask the question: even with a large boost factor,
can neutralino annihilation produce the ``bump'' seen in
the positron spectrum just above 10 GeV?  Even if there 
is a line signal direct from the annihilation, it gets
spread out by the propagation, so that a bump does not get
produced.  Thus, if one looks ``by eye'', one would conclude
that neutralinos cannot produce the data.  However, this
is an inappropriate way to ask the question. One should instead
study the problem statistically to see if one can find
a neutralino model with a good $\chi^2$.  With a combination
of background and annihilation signal, we are able to 
find statistically reasonable fits to the spectral shape
for boost factors above 30.

\section{Supersymmetric model}

We work in the Minimal Supersymmetric Standard Model (MSSM)\@.  In general, the
MSSM has many free parameters, but with some reasonable assumptions we can
reduce the number of parameters to the Higgsino mass parameter $\mu$, the
gaugino mass parameter $M_{2}$, the ratio of the Higgs vacuum expectation
values $\tan \beta$, the mass of the $CP$-odd Higgs boson $m_{A}$, the scalar
mass parameter $m_{0}$ and the trilinear soft SUSY-breaking parameters $A_{b}$
and $A_{t}$ for the third generation.  In particular, we don't impose any
restrictions from supergravity other than gaugino mass unification.  For a more
detailed definition of the parameters and a full set of Feynman rules, see
Refs.~\cite{bg,coann,jephd}.

The lightest stable supersymmetric particle in the Minimal Supersymmetric
Standard Model (MSSM) is most often the lightest of the neutralinos, which are
superpositions of the superpartners of the neutral gauge and Higgs bosons,
\begin{equation}
\tilde{\chi}^0_1 =
N_{11} \tilde{B} + N_{12} \tilde{W}^3 +
N_{13} \tilde{H}^0_1 + N_{14} \tilde{H}^0_2.
\end{equation}

For many values of the MSSM parameter space, the relic density $\Omega_\chi
h^2$ of the (lightest) neutralino is of the right order of magnitude for the
neutralino to constitute at least a part, if not all, of the dark matter in the
Universe (for a review see Ref.~\cite{jkg}).  Here $\Omega_{\chi}$ is the
density in units of the critical density and $h$ is the present Hubble constant
in units of $100$ km s$^{-1}$ Mpc$^{-1}$.  Present observations favor $h =
0.7\pm 0.1$, and a total matter density $\Omega_{M} = 0.3 \pm 0.1$, of which
baryons contribute roughly $\Omega_bh^2\approx0.02$ \cite{cosmparams}.  Thus we
take the range $0.05\le\Omega_\chi h^2\le0.25$ as the cosmologically
interesting region.  This region can be narrowed somewhat if we consider the
results of CMB anisotropy measurements (summarized in e.g. \cite{cmbdata}),
which favor $\Omega_\chi h^2=0.14\pm0.05$.  We are also interested in models
where neutralinos are not the only component of dark matter, so we separately
consider models with arbitrarily small $\Omega_\chi h^2$.

\begin{table}
\begin{tabular}{rrrrrrrr}
Parameter & $\mu$ & $M_{2}$ & $\tan \beta$ & $m_{A}$ & $m_{0}$ &
$A_{b}/m_{0}$ & $A_{t}/m_{0}$ \\
Unit & GeV & GeV & 1 & GeV & GeV & 1 & 1 \\ \hline Min & -50000 &
-50000 & 1.0 & 0        & 100 & -3 & -3 \\
Max & 50000 & 50000 & 60.0 & 10000 & 30000 & 3 & 3 \\ 
\end{tabular}
\caption{The ranges of parameter values used in the MSSM scans of
Refs.~\protect\cite{bg,coann,bub,neutrate,be,beu_pbar,mbg}.}
\label{tab:scans}
\end{table}

As a scan in MSSM parameter space, we have used the database of MSSM models
built in Refs.~\cite{bg,coann,bub,neutrate,be,beu_pbar,mbg}.  The overall
ranges of the seven MSSM parameters are given in Table~\ref{tab:scans}.  While
the ranges are extreme, most interesting models fall in a much more modest
region of parameter space, with the notable exception that very pure Higgsinos,
thus very large $M_2$, can not be ruled out at present.  The database embodies
one--loop corrections for the neutralino and chargino masses as given in
Ref.~\cite{neuloop}, and leading log two--loop radiative corrections for the
Higgs boson masses as given in Ref.~\cite{feynhiggs}.  For all of the MSSM
models in the scan of parameter space, the database contains results for
expected detection rates of the particles in a variety of neutralino dark
matter searches.  The database includes the relic density of neutralinos
$\Omega_{\chi} h^2$.  The relic density calculation in the database is based on
Refs.~\cite{coann,GondoloGelmini} and includes resonant annihilations,
threshold effects, finite widths of unstable particles, all two--body
tree--level annihilation channels of neutralinos, and coannihilation processes
between all neutralinos and charginos.  The database also includes the
supersymmetric correction to the anomalous magnetic moment of the muon
$a_\mu=(g_\mu-2)/2$ which is important for dark matter searches in light of new
data \cite{agsdata} indicating a deviation from the standard model prediction,
as discussed by e.g.\ Ref.~\cite{gm2darkmatter}.  In this paper we will
identify models which have a large contribution, $10\times10^{-10}\le\Delta
a_\mu({\rm SUSY})\le75\times10^{-10}$, to the anomalous magnetic moment of the
muon as being particularly interesting.

We examined each model in the database to see if it is excluded by the most
recent accelerator constraints. The most important of these are the LEP bounds
\cite{pdg2000} on the lightest chargino mass
\begin{equation}
m_{\chi_{1}^{+}} > \left\{
\begin{array}{lcl}
88.4 {\rm ~GeV} & \quad , \quad & | m_{\chi_{1}^{+}} - m_{\chi^{0}_{1}} | > 3
{\rm ~GeV} \\ 67.7 {\rm ~GeV} & \quad , \quad & {\rm otherwise,} \end{array}
\right.
\end{equation}
and on the lightest Higgs boson mass $m_{h}$ (which ranges from 91.5--112 GeV
depending on $\tan\beta$) and the constraints from $b \to s \gamma$ \cite{cleo}
(we use the LO implementation in DarkSUSY \cite{DarkSUSY}).

\section{Positron flux}

We obtain the positron flux from neutralino annihilation in the galactic halo
following Ref.~\cite{be}.  The model is a true diffusion model and assumes that
the diffusion region of tangled galactic magnetic field is an infinite slab.
This approximation is reasonable since most of the positrons are emitted quite
nearby so that the outer radial boundary is unimportant.  Furthermore, energy
losses due to synchrotron radiation and inverse Compton scattering from the
cosmic microwave background and from starlight are included.  This model
roughly agrees with earlier work \cite{kamturner}, though the inclusion of
inverse Compton scattering from starlight is crucial as it doubles the energy
loss rate.

As we will discuss in the following sections, the positron flux from a smooth
galactic halo is too low to explain the positron excess, as has been discussed
previously \cite{be,kanepositron}.  However, any deviations from smoothness
serve to enhance the annihilation signal, as the annihilation rate is
proportional to the neutralino density squared.  However, we must be careful
that in postulating a boost factor, we do not overproduce the other products of
neutralino annihilation, especially antiprotons and gamma rays \cite{clumpy}.
We do have some freedom here, in that the boost factors for positrons,
antiprotons and gamma rays are not necessarily equal, as their propagation
is not the same.  For example, a nearby clump would serve to increase the
positron flux more than it would increase the antiproton or gamma ray fluxes,
as positrons have the shortest range (they have shorter range than the
antiprotons because of their rapid energy loss).  Many of the antiprotons
come from far away, outside the location of the clump, almost a third of
them from as far away as the Galactic Center.  Positrons, on the other
hand, come from much closer, roughly within a few kiloparsecs.

We will fit the full positron dataset of the HEAT experiments (1994 and 1995
combined data \cite{heatfrac} and the 2000 data \cite{newdata}).  We use the
positron fraction data, as the error bars are smaller and the data cleaner.
The full dataset consists of twelve independent measurements of the positron
fraction at various energies.

We will in the following assume that the standard prediction for the positron
background \cite{moskstrong,moskstrongnew} is correct to within a normalization
factor $N$.  We are aware that cosmic ray propagation is not completely
understood, and that even the best efforts to reproduce the observed cosmic ray
spectra need to rely on yet-to-be-understood ad hoc assumptions on the
dependence of the diffusion constant on energy and on the source spectrum
\cite{moskstrongnew}. However, we gather from the latter work that the
discrepancies between the observed and theoretical positron spectra lie
preponderantly at energies smaller than a few GeV, where they can become as
large as a factor of 4 in the hundreds of MeV range. At the slightly higher
energies where the HEAT bump is, the theoretical models
in~\cite{moskstrongnew}, which cover a wide range of theoretical assumptions,
agree to within 20\%.  While this may give some justification to our use of
model 08-005 of Ref.~\cite{moskstrong} as our standard positron background, we
nevertheless stress that it may be possible to explain the positron bump by
purely astrophysical means (although we do not know how).  Keeping these
uncertainties in mind, we proceed with the assumption that the background
calculation is correct, and we study the possibility that neutralino
annihilation can account for the excess positrons.

We assume that the positron signal from neutralino annihilation can be rescaled
by a normalization factor (boost factor) $B_s$.  We find that the best fit
normalization of the background with no signal from neutralinos is $N=1.14$,
with $\chi^2=3.33$ per degree of freedom.  When adding the signal, we make a
simultaneous fit of the normalization of the background $N$ and the
normalization of the signal $B_s$, for each supersymmetric model in the
database.  We say that a given model ``gives a good fit to the positron data''
when: (1) the background-plus-signal fit fits the data better than the
background-only fit with a decrease in $\chi^2$ per degree of freedom greater
than unity, namely the background-plus-signal fit has $\chi^2\le2.33$ per
degree of freedom; (2) the best fit normalization of the background $N$ is
between 0.5 and 2.0, namely the calculation of \cite{moskstrong} is correct to
within a factor of two according to the best fit.

The positron fluxes are more than an order of magnitude smaller than the HEAT
measurements, and we find that the best fit normalizations of the signal $B_s$
lie between 30 and $10^{10}$.  Values of $B_s$ as large as $10^{10}$ are hardly
realistic, but $B_s$ up to 100--1000 might be acceptable given the
uncertainties in the halo structure (the halo could e.g.\ be clumpy
\cite{clumpyhalo,clumpy,moore-etc}).

\subsection{Antiprotons}
In addition, we require that the antiproton flux from annihilations
\cite{pbarhalo,beu_pbar} not be too large, given the boost factor required for
each model.  There is a significant correlation between the antiproton and
positron fluxes due to neutralino annihilations (see e.g.\ Fig.~8 of
Ref.~\cite{be}), so this constraint is important.  Following
Ref.~\cite{clumpy}, we take the antiproton flux as
\begin{equation}
\Phi_{\overline{p}}=k(1+0.75B_s)\Phi^{\rm smooth}_{\overline{p}},
\end{equation}
where $k$ represents the difference in enhancement factors between the
antiprotons and positrons.  The factor $0.75$ comes from the fact that the
antiprotons that reach the Earth on average are produced further away than the
positrons.  In particular, the antiprotons produced close to the galactic
center make a significant contribution to the flux at Earth.  In these denser
environments, a clumpy distribution enhances the signal less than in our local
environment, and hence the different scalings of the positron and antiproton
fluxes (see Ref.~\cite{clumpy} for more details).

We take the constraint from the combined 1995 and 1997 data of the BESS
collaboration \cite{BESSdata}. Note that we use the BESS data rather than the
HEAT 2000 antiproton data (the new instrument measures both positrons and
antiprotons) for the following reason: the observed antiproton flux rises with
momentum to a maximum value and then falls. The spectrum from neutralinos, on
the other hand, is flat with momentum and then cuts off before it reaches the
observed peak.  Hence the strongest constraint on neutralinos comes from lower
energy measurements.  BESS goes to much lower energy than HEAT 2000 and thus
places a stronger constraint.  The BESS collaboration measured the cosmic ray
antiproton flux at low energies to be
\begin{eqnarray}
\Phi_{\overline{p}}(T=400-560\;{\rm MeV})&=&1.27^{+0.37}_{-0.32}\nonumber \\
&\times&10^{-6}(\rm{cm^2\;s\;sr\;GeV})^{-1}.
\end{eqnarray}
Using the central value as the maximum allowed annihilation flux, and taking
$k=1$, we find no models with a boost factor $B_s<100$, though there are a
handful of models with $B_s<300$, including several with a significant value of
$a_\mu$.  Taking $k=0.2$, we find the constraint much less punishing, and given
the uncertainties, we choose this value instead.

\subsection{Two Successful Models}
\begin{figure*}
\centerline{
\epsfig{file=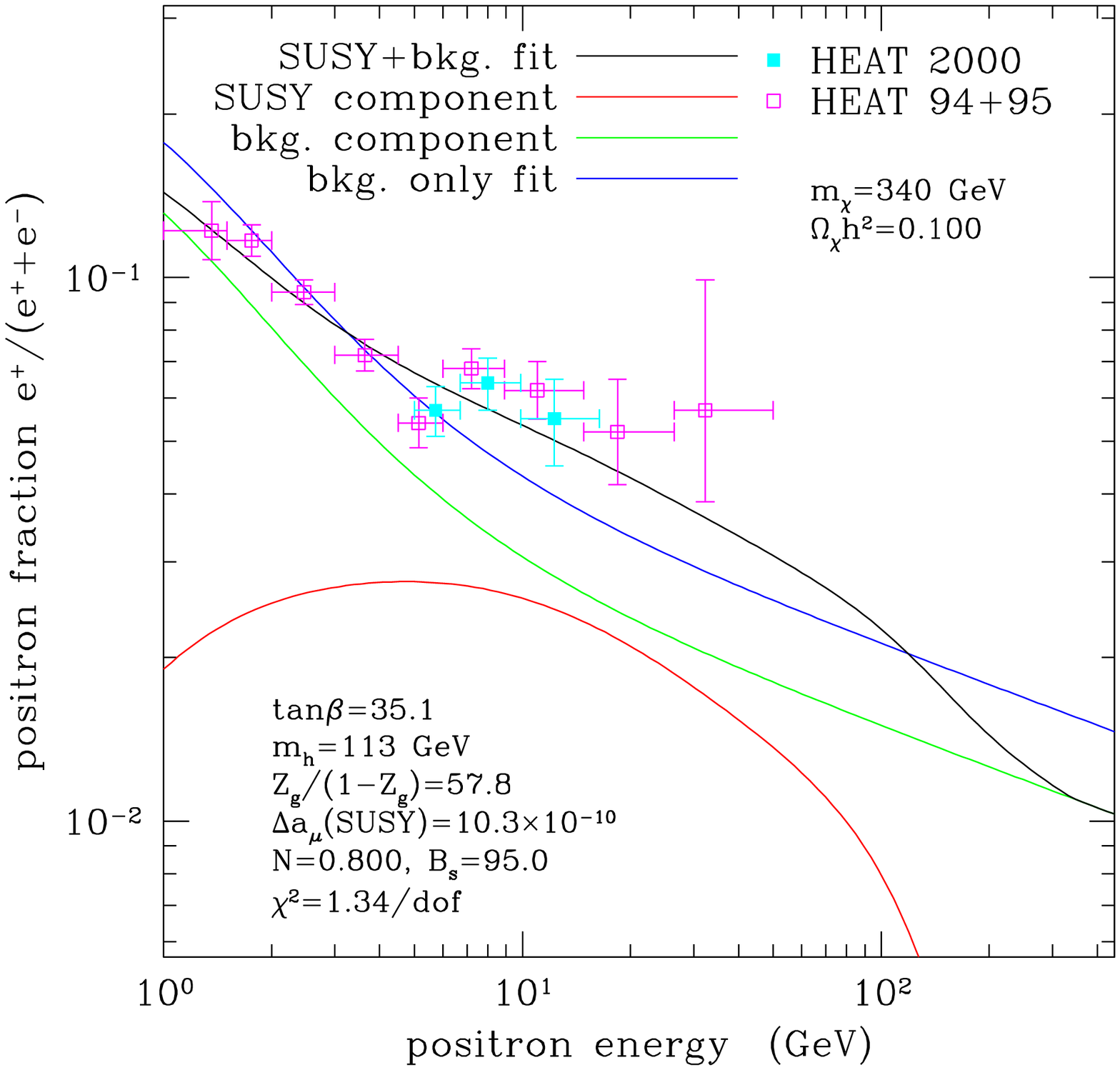,width=0.49\textwidth}
\epsfig{file=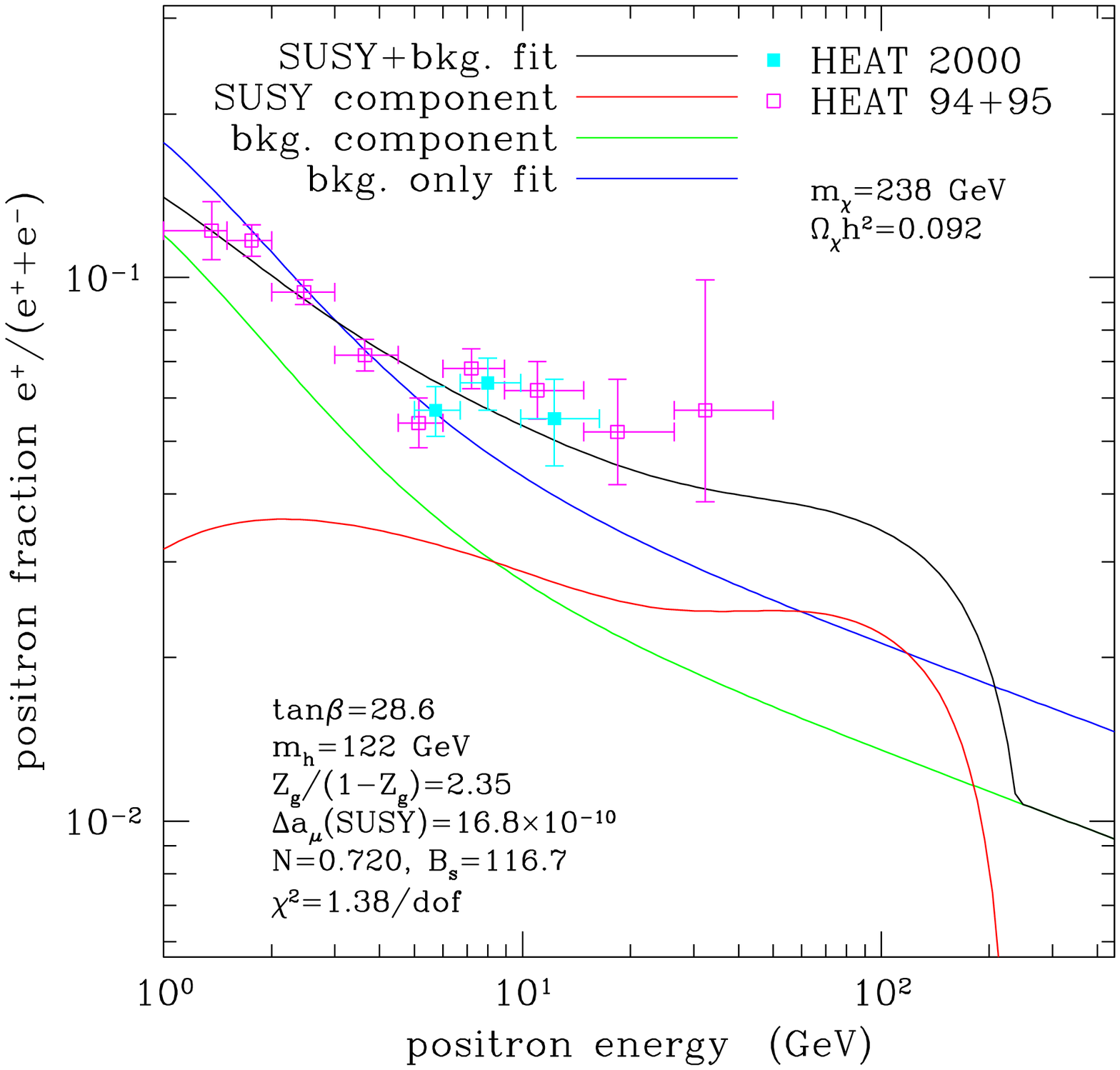,width=0.49\textwidth}}
\caption{Positron fraction data and fits.  We illustrate positron data from
HEAT 94+95 and HEAT 2000, a background only fit, and a SUSY+background fit from
two interesting models from the MSSM database.  Two additional curves
separately display the SUSY and background components of the combined
SUSY+background fit.  These models are gaugino dominated and have contributions
to $a_\mu$ in line with the experimental discrepancy.  The model in Fig.~1a has
positrons primarily from hadronization, while the model in Fig.~1b has hard
positrons from direct gauge boson decays.}
\label{fig:spectrum}
\end{figure*}

In Fig.~\ref{fig:spectrum} we plot the positron data from the HEAT 94+95 and
HEAT 2000 experiments, together with the background only fit, and two
interesting SUSY models that have good fits as well as large contributions to
$a_\mu$.  The antiproton constraints have been applied with $k=0.2$.  Note that
we have found the boost factor to be only weakly dependent on $\tan\beta$, so
that the values of $\tan\beta$ in the figure do not play any important role
(except that $\tan\beta$ is correlated with the SUSY contribution to $a_\mu$).
For other models see Fig.~7 of Ref.~\cite{be}.

The apparent sharp increase in the positron fraction around 7 GeV is not
evident in any of our SUSY models, even before the smoothing effects of energy
loss on the spectrum.  In principle, positrons from direct gauge boson decays
have a perfectly flat spectrum (before propagation) with cutoffs at
\begin{equation}
E_\pm=\frac{m_\chi}{2}\left(1\pm\sqrt{1-\frac{m^2}{m_\chi^2}}\right),
\end{equation}
where $m$ is the gauge boson mass.  For $W^\pm$, a feature at 7 GeV would thus
be had for $m_\chi=238$ GeV.  However there are also positrons from
hadronizations at least from the hadronic gauge boson decays, and possibly from
direct annihilations to quark-antiquark pairs.  The hadronic component is
dominant at the lower cutoff (but not always at the upper cutoff), which means
that we can not reproduce a sharp bump at 7-8 GeV as indicated by the data, but
rather a smoother bump over a larger energy range as seen in
Fig.~\ref{fig:spectrum}.

A way to sharpen the neutralino annihilation positrons into a bump is to have
them all come from a nearby clump which is smaller than the propagation
length. Then a line signal would not be smeared out.  This problem has not yet
been treated in depth.  It requires a different solution of the diffusion
equation and is the subject of a future work.

\section{Favored region in supersymmetric parameter space}

\begin{figure*}
\centerline{
\epsfig{file=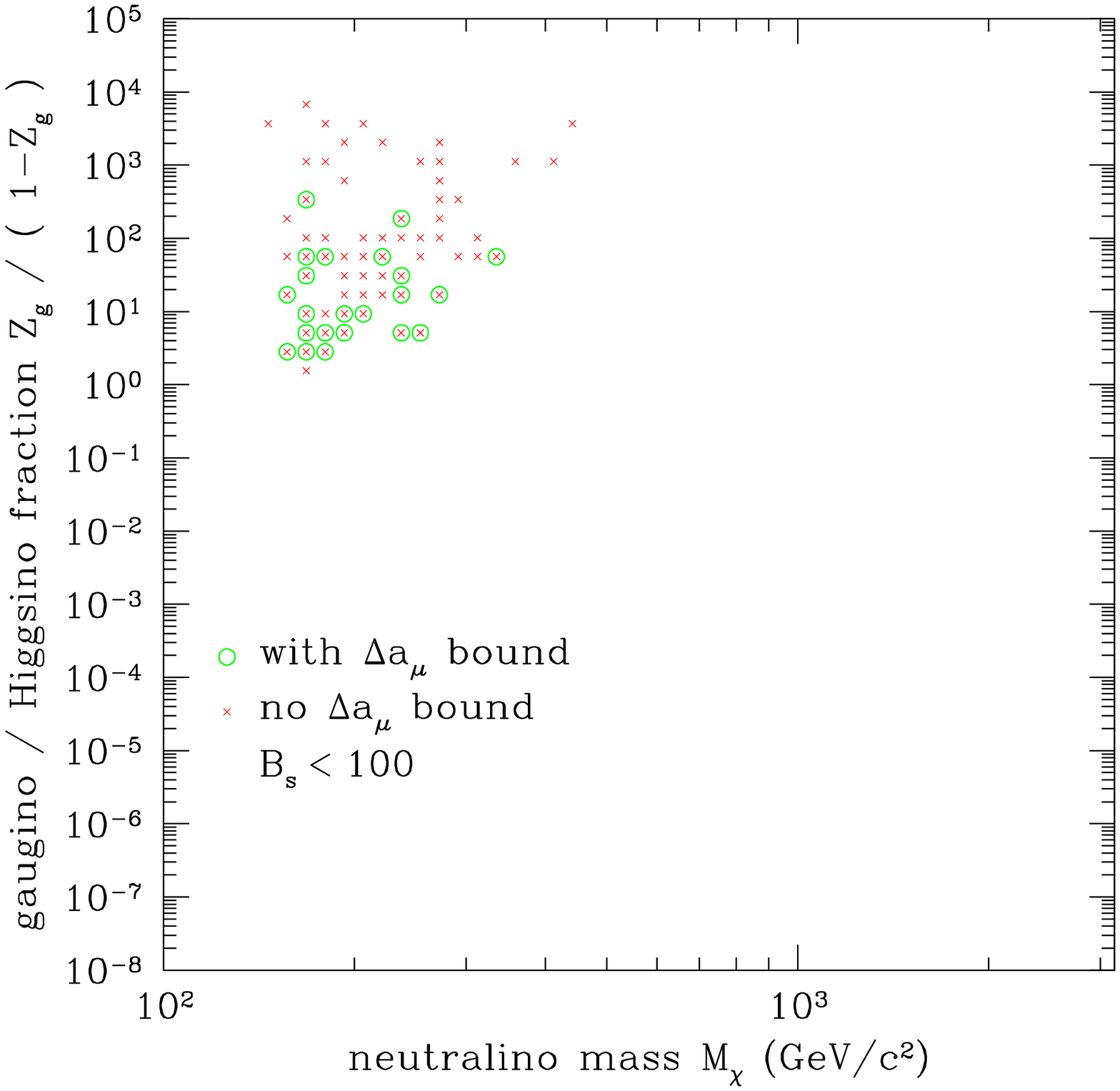,width=0.49\textwidth}
\epsfig{file=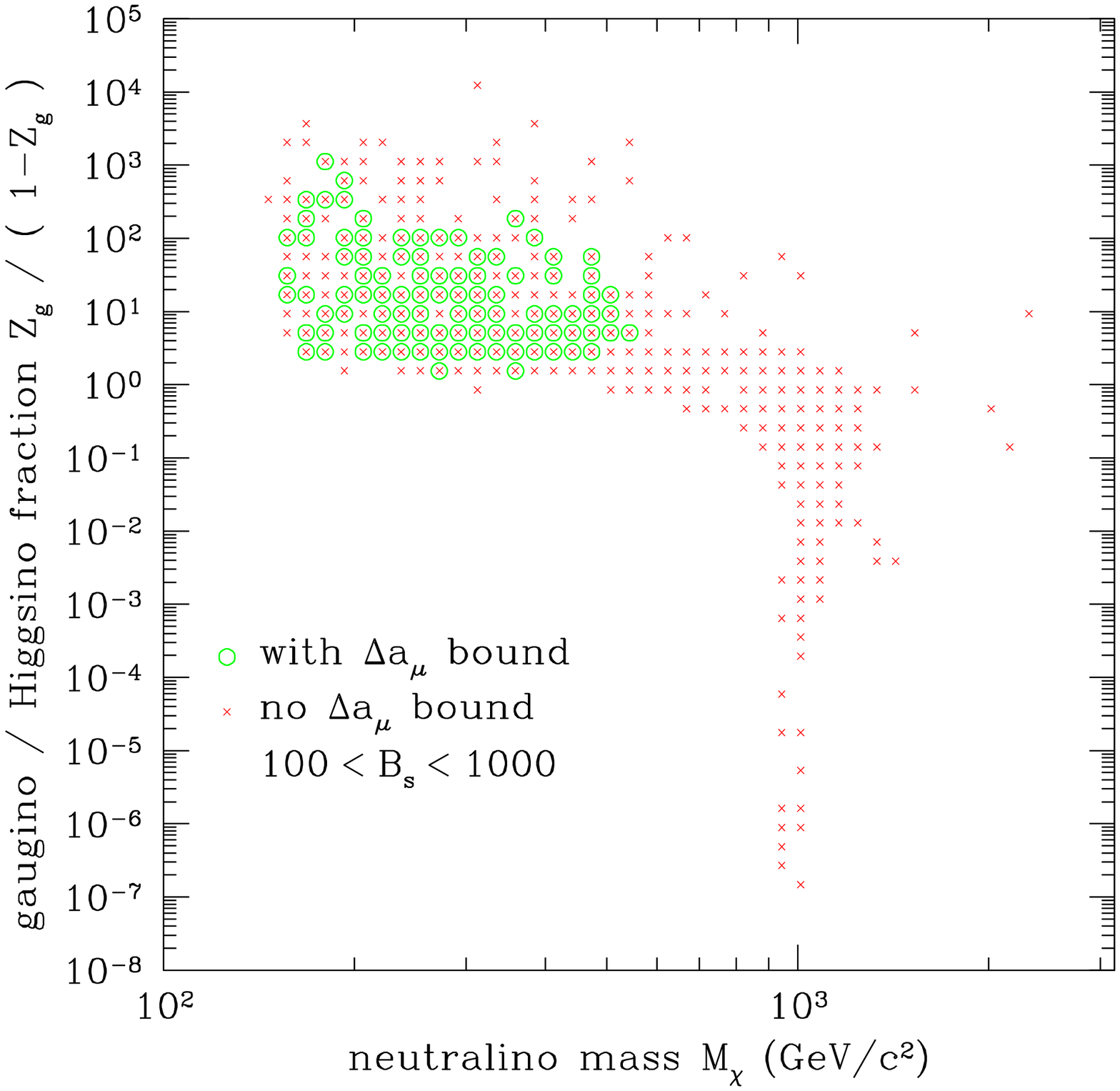,width=0.49\textwidth}}
\caption{Neutralino composition vs.\ mass for well-fitting models.  For
$B_s<100$ these are mixed and gauginos, mostly from $150-400$ GeV.  For
$100<B_s<1000$ the masses extend to 2 TeV, and very pure Higgsinos are also
allowed.  In both cases many of the models have a contribution to $a_\mu$ in
line with the measured discrepancy.}
\label{fig:mxzg}
\end{figure*}

Having computed the positron flux and required enhancement factors to give a
good fit to the positron data, we can now study the supersymmetric parameter
space and identify the favored regions.  The composition of the neutralino,
namely if it is gaugino or Higgsino, is perhaps its most interesting property.
As our indicator of composition, we use the gaugino to Higgsino ratio
\begin{equation}
\frac{Z_g}{1-Z_g}=\frac{|N_{11}|^2+|N_{12}|^2}{|N_{13}|^2+|N_{14}|^2}.
\end{equation}
In Fig.~\ref{fig:mxzg} we plot this ratio vs.\ the neutralino mass separately
for models with $B_s<100$ and $100<B_s<1000$, good $\chi^2$, good background
normalization as discussed in the previous section, and also a relic density in
the region favored by the CMB, $\Omega_\chi h^2=0.14\pm0.05$.  For $B_s<100$,
we find that the models are gaugino--dominated, though some have significant
Higgsino fractions.  The masses of the models are in the range 150--400 GeV for
the most part.  For $100<B_s<1000$, the masses are as large as 2 TeV, and some
very pure Higgsinos become allowed.  For both cases there are a significant
number of models which have a large contribution to $a_\mu$.

\section{Boost factor}

It is instructive to study the best fit boost factor as a function of the
supersymmetric parameters of the models under discussion.  As in the previous
sections, we will restrict ourselves to models that provide a good fit to the
HEAT data, have a relic density in line with CMB data, and do not produce an
overabundance of antiprotons.

\begin{figure*}
\centerline{
\epsfig{file=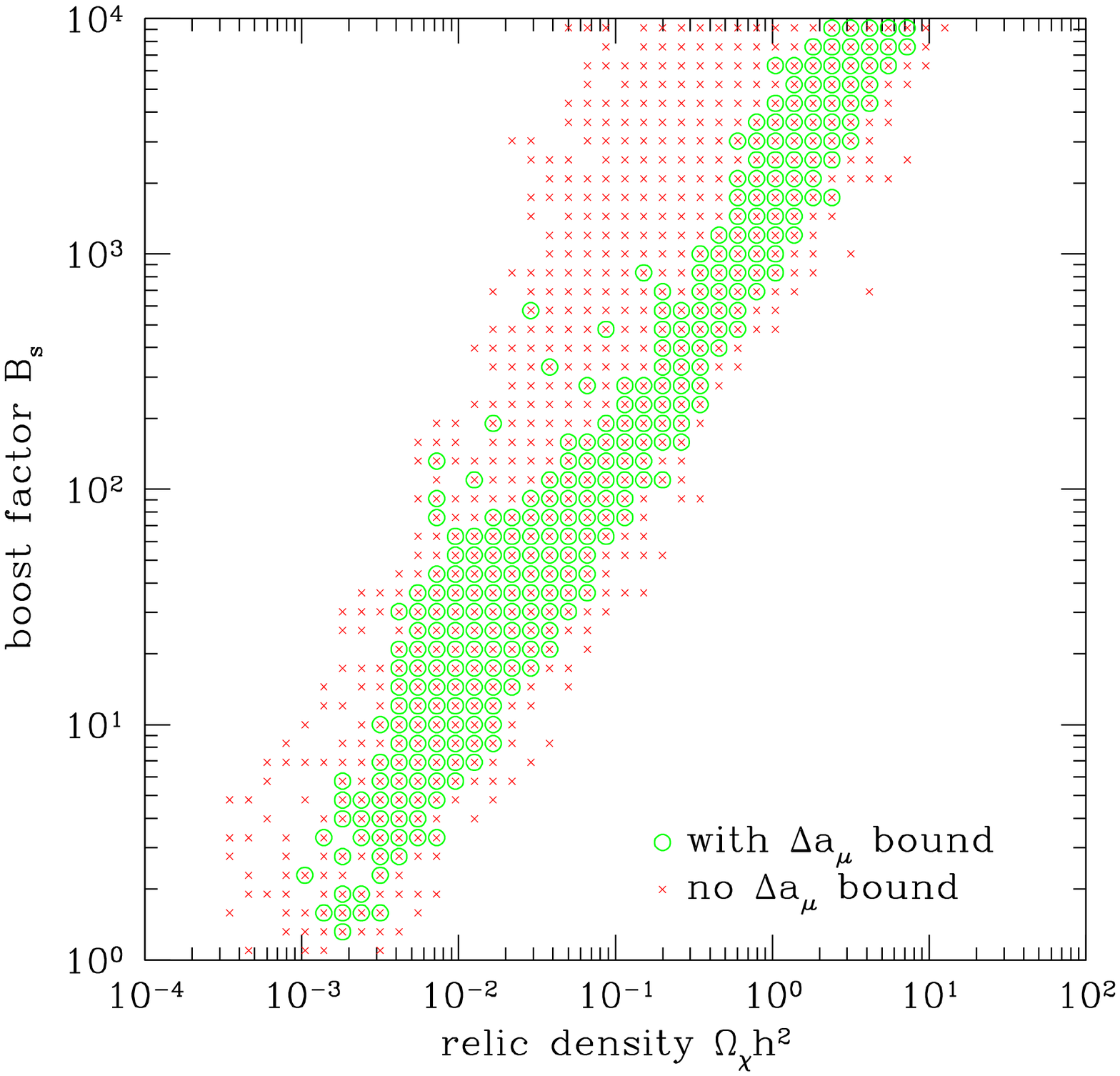,width=0.49\textwidth}
\epsfig{file=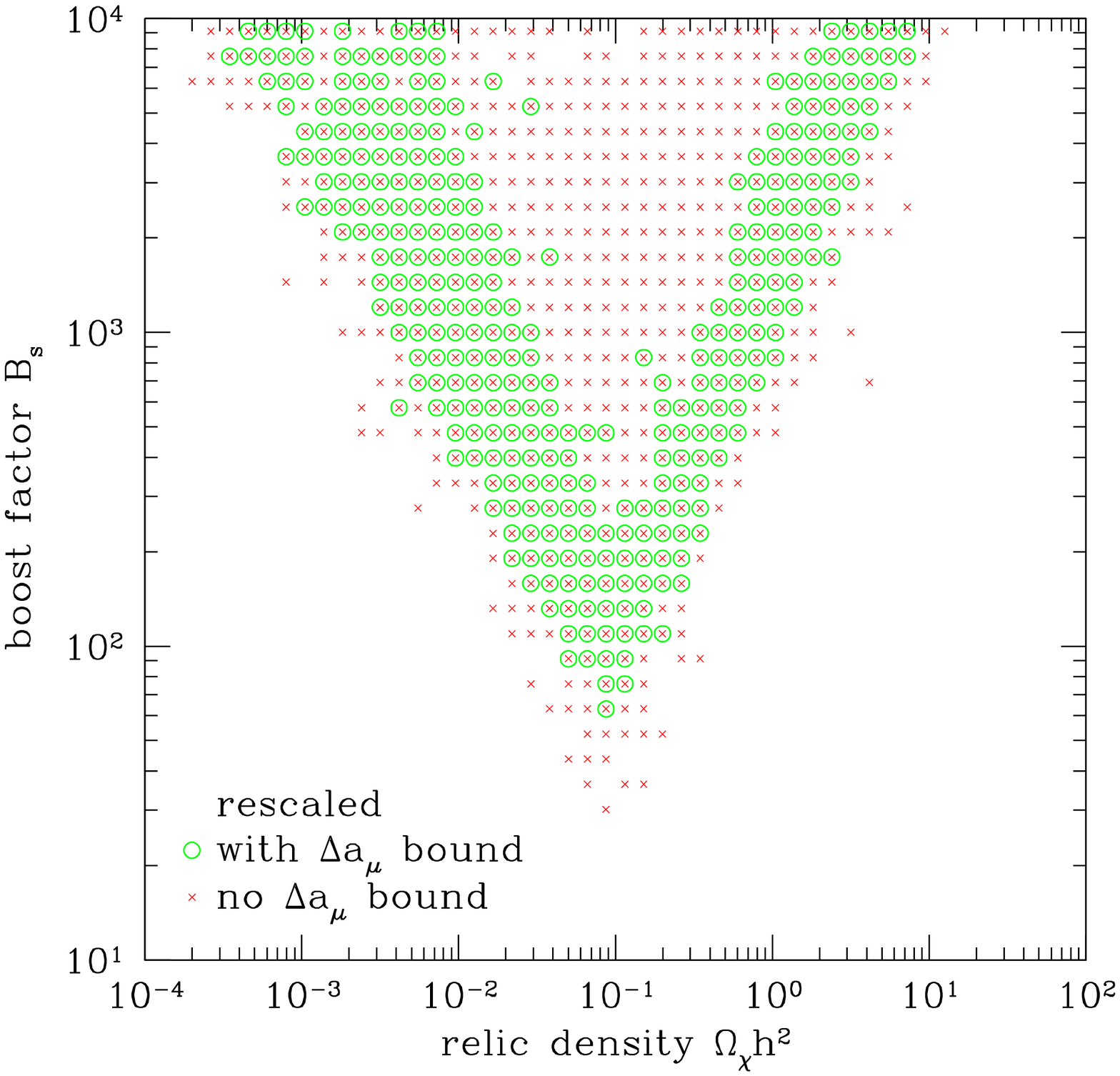,width=0.49\textwidth}}
\caption{Boost factor versus relic density.  In the first panel, no rescaling
is done.  The trend that small boost factor indicates small relic density is
clearly seen.  In the second panel, the rescaling is performed for models
whose relic density is less than $\Omega_\chi h^2 = 0.09$, and it is clear
that the smallest boost factors come from the smallest relic densities that
require no rescaling.}
\label{fig:omega}
\end{figure*}

We first study the dependence of the relic density on the boost factor in these
models.  To do so we will of course neglect to apply the constraint on relic
density, though we retain all other constraints.  We will show two cases, first
assuming that the dark halo density is independent of the relic density, and
second applying a rescaling.  This rescaling is applied for models whose relic
density is less than $\Omega_\chi h^2=0.09$, the lower value in the
CMB range, and is defined as follows.  For low relic densities, 
neutralinos would only make a fraction of
the dark halos of galaxies, and in principle that fraction should be
proportional to the relic density,
\begin{equation}
\rho_{\chi,\rm gal} = \left(\frac{\Omega_\chi}{\Omega_{\rm
CDM}}\right)\,\rho_{\rm CDM,gal}.
\end{equation}
Here, subscript ``gal'' indicates that the density is that inside the Galaxy
and subscript ``CDM'' refers to the dominant matter component.  As annihilation
depends on the square of the density, we rescale as the square of the fraction.
The rescaling affects the best fit boost factor and the antiproton flux as
follows:
\begin{eqnarray}
B_s & \rightarrow & B_s\left(\frac{\Omega_\chi h^2}{0.09}\right)^{-2}, \\
\Phi_{\overline{p}}^{\rm smooth} & \rightarrow & 
\Phi_{\overline{p}}^{\rm smooth}\left(\frac{\Omega_\chi h^2}{0.09}\right)^2.
\end{eqnarray}
In Fig.~\ref{fig:omega}
we plot the boost factor versus the relic density for both cases, not rescaled
and rescaled.  Without rescaling, there is a clear trend that $B_s$ increases
linearly with $\Omega_\chi h^2$. This is expected because the relic density is
inversely proportional to the annihilation cross section, and so is the boost
factor.  When taking the rescaling into account, we find that the lowest boost
factors are required for the models with the smallest relic density without
rescaling, that is with $\Omega h^2 = 0.09$ according to our choice.

We take two main points from Fig.~\ref{fig:omega}.  First, our preferred region
depends somewhat sensitively on the cuts we make in the relic density.
Enlarging our definition of the region of cosmological interest would have a
significant effect on the number of allowed models in the database.  Our
conclusions would however be broadly similar.  Second, and more importantly, we
see the fundamental problem of explaining the positron excess with neutralino
annihilation.  Given the observed value of the dark matter density, the
expected annihilation cross section is too small to explain the observed excess
of positrons without some boost due to clumping or some other mechanism (for
example some models discussed in Refs.~\cite{kanepositron,exotic}).  This point
is independent of the specific model for the WIMP, and only relies on the fact
that the relic density is due to a thermal freeze--out of a stable (or very
long--lived) species, and reasonable annihilation branching fraction to
hadrons.

Concerning the other supersymmetric parameters, we find that the boost factor
is only weakly dependent on $\tan\beta$ and $m_0$, the sfermion mass scale.  We
see a rough trend that models with heavy neutralinos need larger boost factors,
but this is simply related to the fact that the number density scales as the
inverse of the mass and thus the annihilation rate scales as the inverse square
of the mass.

\begin{figure}
\epsfig{width=\columnwidth,file=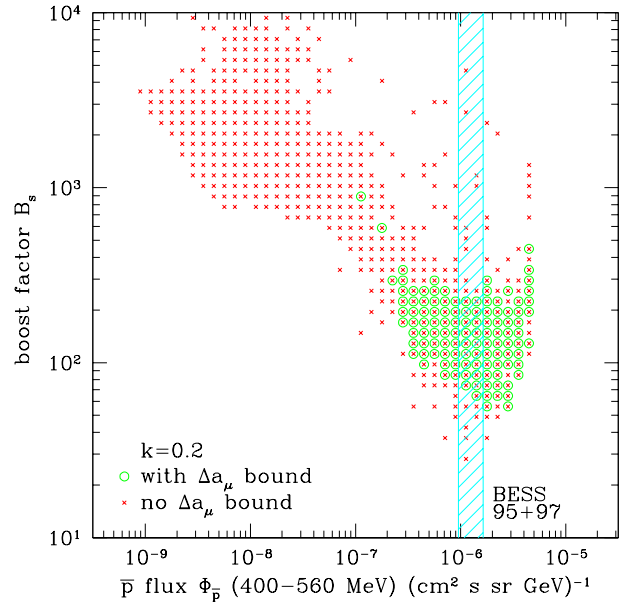}
\caption{Boost factor versus antiproton flux.  The trend that models with small
antiproton fluxes require large boost factors in the positron signal reiterates
the statement that there is a significant correlation between the antiproton
and positron fluxes.  The hatched band indicates the BESS measurement
\protect\cite{BESSdata}.}
\label{fig:pbar}
\end{figure}

Since the constraint from the antiproton flux is so important, we now show how
this quantity depends on the required boost factor.  Of course we now neglect
to apply the constraint on antiproton flux, but we retain all other
constraints.  In Fig.~\ref{fig:pbar} we plot the boost factor vs.\ the
antiproton flux in the 400--560 MeV bin for easy comparison with the BESS
experiment, shown as the hatched band.  That small antiproton fluxes imply
large boost factors is another statement of the fact that the antiproton and
positron fluxes are significantly correlated.  Furthermore, we see the
advantage of allowing $k<1$, as the bound on antiproton flux sits at
$1.27^{+0.37}_{-0.32}\times10^{-6}$ (cm$^2$ s sr GeV)$^{-1}$.  Even for
$k=0.5$, a significant number of models becomes allowed, especially with boost
factors $B_s<300$.

Finally, we comment on the feasibility of the required boost factors, $B_s>30$.
It is well known that dark matter is clumpy in a large range of length scales;
such clumps are clusters, galaxies, dwarf galaxies, etc.  On such large scales
the enhancement factor is in excess of 100 according to simulations of large
scale structure~\cite{moore-etc}.  The question for us is whether such clumps
persist at scales smaller than several kiloparsecs, which is the size of the
emission region for positrons detectable at the Earth.  Unfortunately, there is
really no data at these distance scales, either observational or from
simulations.  Without evidence to the contrary, we must allow such enhancements
to be possible.

In obtaining a boost factor, we have assumed that we can average over a volume
containing many small clumps.  If the halo is not smooth, but we can average
over a large volume (relative to the propagation length of positrons from
several clumps), then we can pull out an enhancement factor.  In other words,
we can use the results of the DarkSUSY code for a smooth halo and just multiply
the result by a boost factor.

\section{Other dark matter searches}

In order to be convinced of an exotic interpretation of cosmic ray data, we
would like confirmation by some other technique.  In this section we discuss
other dark matter search techniques that might give us more confidence that the
positron excess really is due to an exotic primary component.  In particular,
we discuss direct detection of neutralinos by elastic scattering, indirect
detection by gamma ray lines, and furthermore by neutrinos from capture and
annihilations in the centers of the Earth and Sun.

\subsection{Direct detection}

Direct detection of galactic halo neutralinos is one of the most promising
techniques for detecting dark matter, and there are several experimental
collaborations undertaking this program, e.g.\ DAMA \cite{dama}, CDMS
\cite{cdms}, CRESST \cite{cresst}, EDELWEISS \cite{edelweiss}, Cryoarray
\cite{cryoarray}, GENIUS \cite{genius}, IGEX (CanFranc) \cite{IGEX}, HDMS
\cite{HDMS}, MIBETA, ROSEBUD \cite{ROSEBUD}, LiF/TOKYO \cite{tokyo}, UKDMC
\cite{ukdm}, SACLAY, ELEGANT V \cite{elegantV}, and Baksan.  In the next ten
years it is expected that neutralinos with elastic scattering cross sections on
nucleons as low as $10^{-9}$ picobarns or perhaps even lower can be probed
\cite{cryoarray,genius}.  The rates in detectors only depend on the local halo
density at present, so they will not put any severe constraints on the
clumpiness of the halo as a whole.  These rates can of course be enhanced if we
happen to be inside a clump at present \cite{fsw}.

\begin{figure}
\epsfig{file=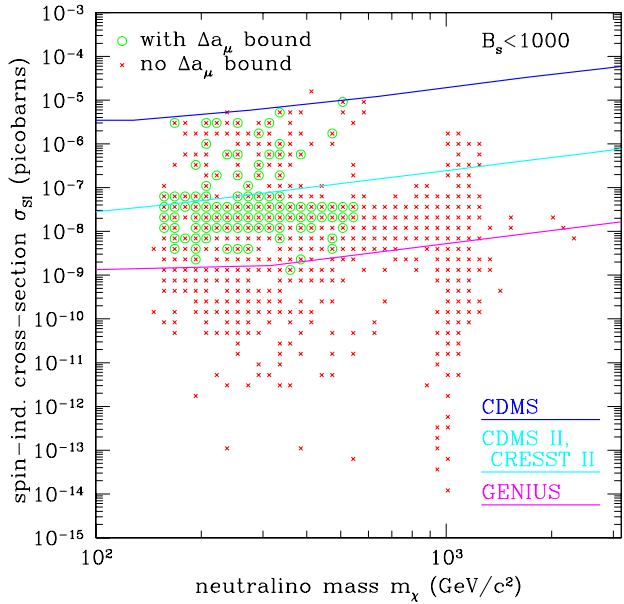,width=\columnwidth}
\caption{Neutralino--nucleon elastic scattering cross section.  We only show
models passing the goodness of fit cuts and require $B_s<1000$.  The current
CDMS exclusion \protect\cite{cdms} (solid line) is plotted along with the
expected reach of the CDMS, CRESST \protect\cite{cresst} (dotted line) and
GENIUS \protect\cite{genius} (dashed line) experiments.}
\label{fig:direct}
\end{figure}

Direct detection is especially exciting in light of the measurement of a
possible discrepancy in the anomalous magnetic moment of the muon
\cite{agsdata}.  Models with large contributions to $a_\mu$ tend to also have
large elastic scattering cross sections \cite{gm2darkmatter}, and we find that
many of the models that can explain the positron excess with boost factors
$B_s<1000$ also have large contributions to $a_\mu$.  We plot the scattering
cross section in Fig.~\ref{fig:direct}.

\subsection{Neutrinos from the Earth and Sun}

Another possible method to detect neutralino dark matter is neutrino
telescopes, such as at Lake Baikal \cite{baikal}, Super-Kamiokande
\cite{superk}, in the Mediterranean \cite{antares}, and the South Pole
\cite{amanda}.  Neutralinos in the galactic halo undergo scatterings into bound
orbits around the Earth and Sun, and subsequently sink to the centers of these
bodies, possibly giving a detectable annihilation signal in neutrinos at GeV
and higher energies \cite{theorists}.  The detectability of this signal is
strongly correlated with the neutralino--nucleon cross section. As discussed in
Ref.\cite{clumpy}, the signal is not likely to be sensitive to the clumpiness
in the halo.  This statement assumes that equilibration time between capture
and annihilation is relatively long and that, averaging over the lifetime of
the Galaxy, the average local density is neither overdense nor underdense.

\subsection{Gamma rays}

Gamma rays from annihilations, both a continuum component \cite{gammahalo} and
a monochromatic component \cite{gammaline} may provide another handle on
neutralinos in the galactic halo.  Experiments such as the GLAST satellite
\cite{glast} and Atmospheric \v Cerenkov Telescopes (ACTs), such as VERITAS
\cite{veritas}, STACEE \cite{stacee}, CANGAROO-III \cite{cangaroo}, and MAGIC
\cite{magic}, may have the necessary sensitivity to detect annihilation photons
in our galaxy above the background \cite{sreekumar}.

To minimize the impact of the halo model and of experimental uncertainties, we
concentrate on the flux at high latitudes, $b > 60^\circ$ and $0^\circ < \ell <
360^\circ$ ($\Delta\Omega = 0.84$~sr), although we also consider the flux
towards the galactic center.  A modified isothermal profile gives
\begin{equation}
J(90^\circ)=0.93(1+1.8B_s),
\end{equation}
where the gamma ray flux is given by
\begin{eqnarray}
\Phi_\gamma&=&1.878\times10^{-13}\left({\rm cm^2\;s\;sr}\right)^{-1}
\times\nonumber \\
&&\frac{N_\gamma\langle\sigma v\rangle}{10^{-29}\;\rm
cm^3\;s^{-1}}\left(\frac{m_\chi}{100\rm\;GeV}\right)^{-2}\,J.
\label{eq:gammas}
\end{eqnarray}
There is only a very weak halo model dependence in this result for $J$ at high
galactic latitude \cite{clumpy}.  We might exclude models which have too high a
gamma ray flux as compared with the measured value at high latitude
\cite{sreekumar},
\begin{equation}
\Phi_\gamma(E>1\;{\rm GeV})=(1.0\pm0.2)\times10^{-6}\rm(cm^2\;s\;sr)^{-1},
\end{equation}
though with boost factors $B_s<100$ the antiprotons are always more powerful
\cite{clumpy}.  However, boosting the signal of the gamma ray lines may allow
their detection, which would be a clear confirmation of the neutralino halo.

The sensitivities of gamma-ray detectors 
to the gamma ray lines can be computed following
Ref.~\cite{bub}.  First the exposure is determined as a function of energy, as
the ACTs in particular have an effective collection area that depends on
energy.  For the ACTs we consider, these are of order $10^8$ cm$^2$ near
threshold and rising to 10$^9$ and more at TeV energies.  ACT integration times
of 500 hours are assumed, while a 2 year GLAST integration is assumed.  The
GLAST exposure is taken to be a constant 1800~cm$^2$~sr, which simply
multiplies the 2 yr integration, the fraction of time pointing towards a target
already accounted for.  The angular field of view for the ACTs is taken to be
0.01 sr, a circle 3.5$^\circ$ in radius.  Based on these exposures, the number
of background events is determined from the extragalactic gamma ray background,
and additionally for ACTs, the backgrounds of cosmic ray electrons and
misidentified hadrons.  In fact the photon background is unimportant for ACTs:
\begin{eqnarray}
\Phi_{\rm had}&=&1.0\times10^{-2}\left(\frac{E}{\rm
GeV}\right)^{-2.7}\left(\rm cm^2\;s\;sr\;GeV\right)^{-1},\\
\Phi_{e^-}&=&6.9\times10^{-2}\left(\frac{E}{\rm
GeV}\right)^{-3.3}\left(\rm cm^2\;s\;sr\;GeV\right)^{-1},\\
\Phi_{\gamma}&=&6.0\times10^{-5}\left(\frac{E}{\rm
GeV}\right)^{-2.7}\left(\rm cm^2\;s\;sr\;GeV\right)^{-1}.
\end{eqnarray}
We note that the background flux for a gamma ray line at a specific energy need
only be integrated over the energy resolutions of the experiments, taken to be
fractionally 0.15 for ACTs and 0.015 for GLAST.

The signal is obtained from Eq.~(\ref{eq:gammas}), taking care to properly
average $J$ over the field of view.  For high galactic latitudes and for ACTs
towards the galactic center this is a minor consideration as $J$ changes little
over the field of view, though for GLAST we find that at the galactic center
the averaged value is about ten times smaller than the central value.  We
require a $5\sigma$ excess above background to claim a detection.

\begin{figure}
\epsfig{file=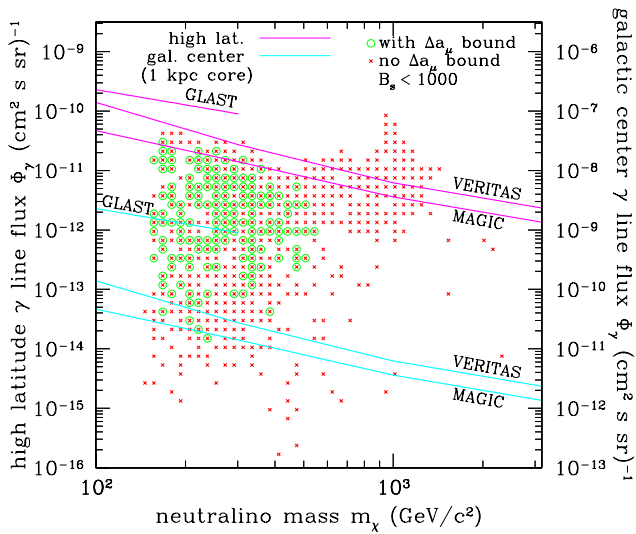,width=\columnwidth}
\caption{Gamma ray line flux.  We include the expected sensitivity (solid
lines) of the VERITAS \protect\cite{veritas}, MAGIC \protect\cite{magic} and
GLAST \protect\cite{glast} experiments.  The left axis is the flux from high
galactic latitude (dotted sensitivity curves), and the right axis is the flux
from the galactic center for an isothermal halo with a 1 kiloparsec core (solid
sensitivity curves).  We show only models passing the goodness of fit cuts and
require $B_s<1000$.}
\label{fig:gammas}
\end{figure}

In Fig.~\ref{fig:gammas} we plot the flux from the gamma ray lines at high
latitude for models with $B_s<1000$, appropriately boosted by $B_s$.  We
include the expected sensitivity of the VERITAS \cite{veritas} and MAGIC
\cite{magic} ACTs as well as the GLAST \cite{glast} satellite.  Furthermore, we
include a prediction appropriate to the galactic center, assuming an isothermal
halo with a 1 kiloparsec core, which has a signal roughly 1000 times larger,
$J\sim 1600 B_s$ (this value is decreased by a factor of 10 when averaged over
the GLAST field of view).  At high latitude, a more sensitive experiment is
probably required, though towards the galactic center, many models would give
detectable fluxes in the gamma ray lines.

\section{Conclusions}

The cosmic ray positron excess is intriguing, as there is no simple
astrophysical model that can explain it.  We are left to consider a primary
component, such as from neutralino annihilations.  We summarize here our
conclusions concerning this scenario.

First, the observed value of the dark matter density implies (assuming thermal
production) an annihilation cross section that is too small to reproduce the
positron excess without some form of enhancement.  This is a general statement,
not tied to a specific model.  We thus resort to enhancing the signal;
fortunately such an enhancement is natural as the dark halo is expected to be
clumpy.  This leads to a second difficulty, namely that one can not enhance the
positron signal without enhancing other signals, especially antiprotons which
would also be produced by annihilation.  Hence antiprotons provide a further
constraint on this scenario.  Indeed lowering the antiproton flux is a further,
albeit small, price to pay in the neutralino annihilation scenario. (Note that
antiprotons come from much farther away than the positrons, so their fluxes are
not always directly correlated.)  In addition, in order to obtain a positron
spectrum that matches all the data, we had to adjust the normalization of the
background (another price we had to pay).  We had to choose a positron
background a factor of 2 lower than the standard fit to the positron data with
background alone. The reason for this lowered background normalization is that
one cannot overshoot the data at energies 1-3 GeV.  In reality the background
is not terribly well understood, and though it cannot by itself explain the
turnup in the data at 10 GeV, one wonders if perhaps the boost factor might not
be plausibly lower than the values we have found.  However, we find this
possibility unlikely.  We should mention here that the propagation
uncertainties make the change in background normalizations and relative boosts
between antiprotons and positrons (the $k=0.2$ vs.\ $k=1$ issue) more plausible,
and we do not believe these to be serious concerns with our analysis.

Second, assuming that the boost factor is between 30 and 100, we find
gaugino--dominated SUSY models that satisfy all constraints, have neutralino
masses in the 150--400 GeV range, and have a large contribution to the
anomalous muon magnetic moment $a_\mu$.  Allowing boost
factors as large as 1000 extends the mass range to 2 TeV, and furthermore
allows Higgsino--dominated neutralinos.  Such boost factors are certainly
plausible, and with no evidence to the contrary, we must take this possibility
seriously.

Confirmation of the annihilation hypothesis could come from several approaches.
The direct detection of halo neutralinos would certainly be a powerful
indicator, as would neutrinos from the center of the Earth and Sun.
Antiprotons and gamma rays could help study the clumpiness of the galactic
halo, helping to determine if it is in fact as large as in the scenario we have
presented.  In particular, boosting the intensity of gamma ray lines may allow
their detection, which would be a clear confirmation of a neutralino halo.  The
next several years will be an exciting time for particle dark matter searches.

\acknowledgments

We thank Greg Tarl\'e and Andy Tomasch for useful conversations and for
providing us with the HEAT 2000 data.  JE thanks the Swedish Research Council
for support.  KF was supported at the University of Michigan in part by a grant
from DOE.  KF and PG wish to thank the Max Planck Institut f\"ur Physik in
Munich, Germany, for hospitality.


\end{document}